# All-Dielectric Structural Coloration Empowered by Bound States in the Continuum

*Hong Zheng[1,2], Haiyang Hu[2], Thomas Weber[2], Juan Wang[2], Lin Nan[2], Bingsuo Zou[1,3], Stefan A. Maier[4,5] and Andreas Tittl[2\*]*

1. Beijing Key Laboratory of Nanophotonics and Ultrafine Optoelectronic Systems, School of Physics, Beijing Institute of Technology, Beijing 100081, China

2. Chair in Hybrid Nanosystems, Nanoinstitute Munich, Faculty of Physics, Ludwig-Maximilians-Universität München, 80539 München, Germany

3. Guangxi Key Lab of Processing for Nonferrous Metals and Featured Materials, School of Resources, Environments and Materials, Guangxi University, Nanning 530004, China

4. School of Physics and Astronomy, Monash University, Clayton, Victoria 3800, Australia.

5. Department of Physics, Imperial College London, London SW72AZ, UK.

\* E-mail: Andreas.Tittl@physik.uni-muenchen.de

KEYWORDS: color printing, bound states in the continuum, titanium dioxide, metasurface

ABSTRACT:

The technological requirements of low-power and high-fidelity color displays have been instrumental in driving research into advanced coloration technologies. At the forefront of these developments is the implementation of dye-free coloration techniques, which overcome previous constraints related to insufficient resolution and color fading. In this context, resonant dielectric nanostructures have emerged as a promising paradigm, showing great potential for high efficiency, remarkably high color saturation, wide gamut palette, and realistic image reproduction. However, they still face limitations related to color accuracy, purity, and simultaneous brightness tunability. Here, we demonstrate an all-dielectric metasurface empowered by photonic bound states in the continuum (BICs), which supports sharp resonances throughout the visible spectral range, ideally suited for producing a wide range of structural colors. The metasurface design consists of titanium dioxide ($TiO_2$) ellipses with carefully controlled sizes and geometrical asymmetry, allowing versatile and on-demand variation of the brightness and hue of the output colors, respectively.

INTRODUCTION

Generating vibrant and well-defined colors has been a motivating factor for the advancement of optics extensively studied since ancient times[1–3]. Dyes and pigments are the predominant materials utilized for color rendering in various applications, selectively absorbing certain visible spectra and reflecting the remaining parts, resulting in reduced brightness and a limited gamut[4]. However, further developments of this passive coloration approach are restricted by wavelength-limited resolution, color fading over time, and vulnerability to photobleaching upon exposure to intense ultraviolet radiation[5]. To address the above concerns, structural coloration, stemming from the precise manipulation of light-matter interaction, has become an alternative approach to tackle the obstacles of dyes and pigments[6,7]. Different from pigment- or dye-based methodologies, such coloration techniques provide several advantages, including superior chromaticity and color gamut, improved spatial resolution, long-lasting coloration, and manufacturing scalability[8-10]. With unparalleled advancements in nanofabrication technology, plasmonic resonances of metallic nanostructure have been intensively studied to demonstrate high-resolution color printing beyond the optical diffraction limit, precisely manipulating the geometry of nanostructures to tailor the reflectance and transmittance spectra[11–15]. However, the plasmonic nanostructures' metallic loss impedes the generation of sufficiently brilliant coloration in the visible spectrum[8].

Alternatively, high refractive index dielectric nanostructures provide an effective way to overcome these challenges[16-24]. As of late, all-dielectric metasurfaces have emerged as a promising technology to produce vibrant colors by exploiting the Mie resonances of individual nanostructures. For instance, titanium dioxide ($TiO_2$) nanostructures, possessing electric and magnetic dipole resonances, have proven to be highly tunable[21] and possess high color saturation[17]. Similarly, silicon nitride ($Si_3N_4$) metasurfaces based on Rayleigh anomalies were employed to suppress high-order Mie resonances at relatively short wavelengths and generate vivid colors[23]. Additionally, color saturation has been improved even further by employing multilayer dielectric stacked nanostructures[19] and matching refractive index layers[22]. However, the above-mentioned metasurfaces lack the ability to easily adjust color brightness, preventing them from creating shadow rendering effects that would enhance spatial and stereo perception. Recently, crystalline silicon metasurfaces can facilitate tailored hue, saturation, and color intensity modulation[20]. Notwithstanding, the implementation of a spatial multiplexing strategy significantly deteriorates linewidth, and high aspect ratio surface features pose nanofabrication challenges.

All-dielectric metasurfaces supporting photonic bound states in the continuum (BIC) have gained considerable attention for their potential in nanoscale lasing[25-29], chiral optics[30-33], nonlinear photonics[34-37], biomolecular sensing[38-41], and photocatalysis[42] due to their exceptional spectral selectivity, strong light confinement, and significant enhancement of electric fields. A true BIC, originally described in quantum mechanics, is a mathematical object with an infinite Q factor and vanishing resonance width in an open system, which cannot couple to any radiation channel propagating outside the system[43-46]. Slight perturbations can break the pure BIC condition and create quasi-BIC (q-BIC) modes, which can be excited and observed from the far field[47]. Symmetry-protected BICs have received growing attention among the various types of BICs, due to their high optical signal contrasts, experimental robustness, and straightforward measurements using brightfield microscopy[38,41,48-51]. Specifically, breaking the in-plane inversion symmetry of the BIC unit cell geometry induces energy dissipation from the resonator system through the radiative loss channel, allowing such resonances to be observed from the far field in reflectance/transmittance spectra. The leveraging of q-BIC modes provides an effective approach to constructing the ideal saturated color pixel. The spectral selectivity provided by the BICs, which exhibit ideal reflectance intensity in the desired wavelength range and near zero background reflectance spectrum outside of that range, align well with the characteristics of "Schrödinger's pixel" based on the considerations of Erwin Schrödinger[52-54]. Schrödinger's methodology exhibits a flat profile of the reflectance spectrum featuring an abrupt shift from zero to unity, resulting in bright and saturated colors. Notably, achieving a Schrödinger's pixel with high saturation and brightness demands both a high reflectance of the intended wavelengths and a simultaneous near-complete reduction in reflectance or scattering across other wavelengths, which holds the potential for a wide range of applications in displays and imaging. Previously, a Schrödinger's red pixel was realized by optimizing the geometry of a q-BIC antenna using a gradient descent approach, which achieves maximum red color saturation while minimizing reflectance at blue and green wavelengths[55]. Yet the considerable intrinsic losses in amorphous Si towards lower wavelengths have impeded efforts to achieve highly saturated blue and green pixels using q-BIC resonance. Furthermore, symmetry-protected BICs can be fine-tuned by modifying their geometrical asymmetry, resulting in changes in their intensity and enabling tunable brightness, which has so far been overlooked. Besides, liquid crystal metasurfaces' color brightness control is hindered by complex architecture and incomplete spectral response for full-color displays[56]. Thus, it continues to be a substantial hurdle to achieve a metasurface platform simultaneously possessing a very narrow linewidth and negligible background spectrum, allowing for the creation of brightness tunability, highly efficient and saturated structural colors across the entire visible spectral range.

Here, we demonstrate a $TiO_2$ metasurface supporting symmetry-protected BICs, which achieves high reflectance, pure color rendition, a wide gamut palette, and photorealistic reproduction. Explicitly, the dielectric metasurface, composed of a zig-zag array of $TiO_2$ resonators on a glass substrate, exhibits ultrasharp reflectance spectra with a low background spectrum, thus providing saturated and bright reflectance spectra in the visible region. Fine-tuning of the geometric asymmetry factor allows for precise control over the radiative loss of the constituent unit cell, capable of tailoring the reflectance (i.e., the color brightness). In simulations, the geometry of the q-BIC antenna is optimized to realize the brightest and most saturated color via tailoring the radiative decay rate. Adjusting the color intensity has the potential to create a shadow-rendering effect that enhances an image's spatial and stereo perception. Experimentally, we obtain color saturation exceeding 162% of the sRGB gamut with maximal ~80% reflectance. As proof-of-concept demonstrations, we further fabricated several metasurfaces with photorealistic impressions.

RESULTS AND DISCUSSOION

**Design of BIC-driven optical metasurfaces**

Vibrant and brightness-tunable colors are generated in reflection by engineering the radiative decay rate of the BIC-driven resonances, achieved by manipulating the asymmetry of the unit cell geometry. The BIC-driven optical metasurface contains two elliptical $TiO_2$ nanostructures on top of a glass substrate, where the main ellipse axes are tilted towards the y-axis to produce a zig-zag array (**Figure 1a**). The chosen two ellipse BIC unit cell geometry enables precise photonic control, resulting in high signal modulation, low spectral background, and fabrication robustness[38-40,55]. The structure height (h) is 120 nm, while the ellipse major-axis ($A_0$) and minor-axis ($B_0$) are 300 and 100 nm, respectively. The pitch along the x and y direction of the 2D periodically arranged nanostructures to form metasurface pixels is 330 nm. The asymmetry parameter is the tilting angle θ between the y-axis and the bar's long axis. The q-BIC nanoantennas are composed of amorphous

TiO$_2$, with n and k values as shown in Figure S1. The optical constants of TiO$_2$ were obtained from ellipsometry. The subsequent scaling and experimental design are all based on these parameters.

To gauge the color performance of this metasurface, we first numerically investigated the reflectance spectra under incident x-polarized light using a finite-element frequency-domain Maxwell solver (for details see Methods). Originating in BIC-inspired physics, the reflection spectrum shows a sharp resonance, where the radiative contribution exhibits the expected characteristic inverse square law with asymmetry. Thus, under normally incident x-polarized light excitation, sharp reflectance with a low spectral background is controlled by the asymmetry parameter introduced by the tilting angle (**Figure 1b**). Specifically, a symmetry-protected BIC structure ($\theta=0°$) would be inaccessible for electromagnetic waves from the far field. As the symmetry is broken and the structure becomes asymmetric ($\theta > 0$), the reflectance spectrum will have higher intensity due to the increased radiative decay rate. Equations for the far-field reflectance and the relationship between the radiative decay rate and the asymmetry factor are given in Supporting Note 1. Notably, the optical response associated with a 20° tilting angle can readily achieve a high saturation and reflectance when the geometric dimensions of the nanoantenna are carefully optimized. Similar to the situation of Schrödinger's pixel, the nanostructure with a tilting angle of 20° can deliver an ultrasharp resonance with the highest intensity and low spectral background, which is ideal for highly saturated color. Alternatively, the designed metasurface exhibits a single sharp reflectance peak in the optical spectrum and highly suppresses the reflectance at other off-resonant wavelengths via engineering the radiative loss rate via the asymmetry factor of specific BIC unit cell, which creates a pronounced increase of the color saturation and gamut. In addition, this architecture allows for straightforward resonance tuning via multiplying the unit-cell dimensions by a scaling factor *S* varying from 0.8 to 1.4, as given by A= $S·A_0$, B= $S·B_0$, P=$S·P_x$. Linear tuning of the resonance position can cover the entire visible region (**Figure 1c**). As the scaling factor increases, the reflectance peak position redshifts from 434 to 702 nm. The dependence of the simulated reflectance spectra for the proposed nanostructure on different structural parameters was also investigated (Figure S2). In general, if the parameter values (h, $A_0$, $B_0$) increase, the resonance background will be higher, or a new higher-order mode will be generated, resulting in less pure color. Conversely, decreasing the corresponding parameter values will cause lower resonance intensity, giving rise to dim colors and a narrow gamut. Remarkably, by optimizing the periodicities of the rectangular array of nanostructures, the narrowest FWHM of the reflectance spectra could reach sub-5 nm and produce an even cleaner resonance background (Figure S3).

After demonstrating the realization of highly saturated colors, our focus shifts to regulating the intensity, i.e., the color brightness. As shown in **Figure 1d**, our dielectric metasurface can be spectrally tuned in intensity via asymmetry parameter θ. The reflectance spectra exhibit redshifts of 20 nm when the pair is changed from 20° to 0°. To enable independent control of color brightness, we fine-tune the scaling factor of the proposed BIC metasurface at different tilting angles using linear tunability of resonance positions, causing the spectral profiles of blue color to remain with only a decreasing magnitude. Through subsequent fine-tuning, the reflectance spectra show continuous changes in color brightness while maintaining the same hue and saturation. Insets show the calculated colors of the nanoantennas based on their reflectance spectra[57] A similar analysis for adjusting the brightness of the green and red pixels is shown in Figure S4, highlighting the potential of our all-dielectric metasurface approach for color filtering devices. Moreover, Figure S5 shows the reflectance spectra of blue, green and red color pixels with different tilting angles of the nanostructures varying from 0 to 20° after a scaling sweep. The RGB pixel spectral profiles remained, but with reduced brightness, preserving hue and saturation.

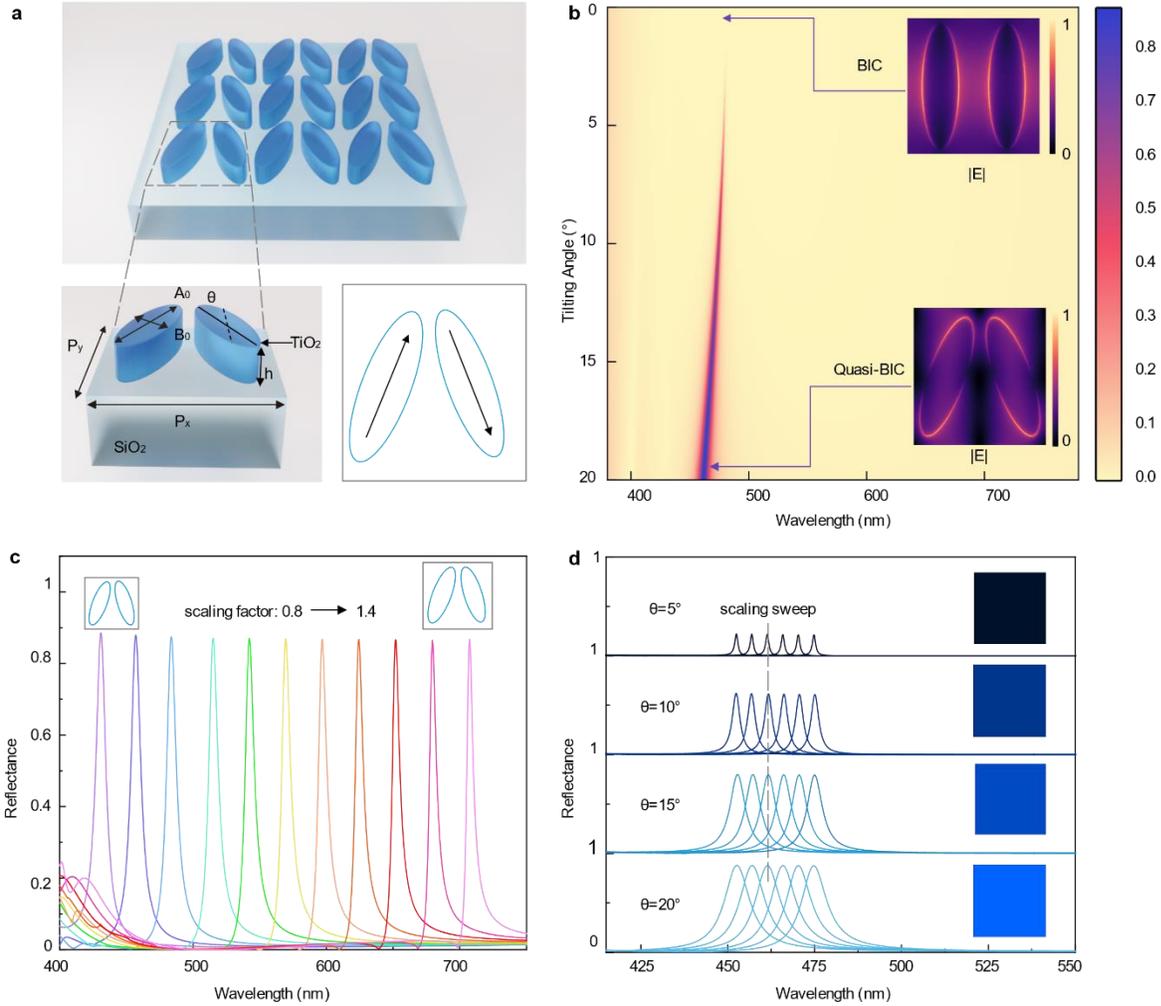

**Figure 1. Numerical analysis of symmetry-protected BICs for the distinct colors and continuous intensity tunability.** (a) The schematic diagram of the $TiO_2$ metasurface supports symmetry-protected BIC and shows the BIC working principle of opposing dipoles. (b) The reflectance spectra are analyzed concerning wavelength and various tilting angles of the unit cell, while keeping the scaling factor S fixed at 0.8. The structure height (h) is 120 nm. The inset shows the electric near-field distribution for corresponding BIC and quasi-BIC. (c) The simulated

reflectance spectra of BIC resonances, adjusted by varying geometric parameters, encompass the entire visible spectrum. (d) Simulated reflectance spectra of BICs with different scaling factors S and asymmetries θ, respectively. The insets show the corresponding calculated colors.

**Fabrication and experimental validation for achieving high color saturation**

The $TiO_2$-driven structural coloration metasurface was fabricated via high-resolution electron beam lithography on a glass substrate. The comprehensive fabrication process is depicted in Figure S6 and described in detail in the Methods. **Figure 2a** shows a brightly full-color photograph of a dielectric chip with an 18×18 optical array, obtained by nanostructuring a 120 nm thick $TiO_2$ film on a glass substrate, and every metasurface has an approximate diameter of 60 μm. Scanning electron microscopy (SEM) images of the fabricated metasurfaces, featuring an array of symmetry-broken nanoresonators tilted at 20°, demonstrate the accurate and uniform reproduction of each nano ellipse's geometry (**Figure 2b**). The minimum gap size between resonators is about 12 nm (**Figure 2c**). While deviating from the numerically predicted optimum values, we selected asymmetry parameters that were deemed appropriate for achieving the desired results, as they offered enhanced reproducibility and stability within the constraints of our nanofabrication processes. **Figure 2d** presents the experimentally measured spectra, which closely resemble the simulated results with scaling factor $S$ varying from 0.84 to 1.4, where the inset shows the simulated color. The experimental colors are associated with their corresponding hue values, as depicted in Figure S7. Significantly, the metasurface spectrum exhibits sharp peaks at given design wavelengths without an additional resonance background. As the scaling factor increases, the high-reflectance spectral region spans from 446 to 704 nm. Both the peak positions and the FWHMs match the experimental results well. Some slight discrepancies between the measured and simulated reflectance spectra result from fabrication variations, measurement errors, and the optical properties of the nanostructure. The reflectance peak decreases to 0.35 within the blue region is attributed to the increased absorption by amorphous $TiO_2$[58]. Additionally, to improve agreement with the simulated CIE map, normalized values below zero have been omitted, with the uncorrected dataset available in Figure S8. We also investigate the influence of incidence polarization on the achieved color saturation (Figure S9). The details of the optical measurements are presented in the Methods section.

To gain a deeper understanding of the highly saturated colors and broad spectral coverage, we computed the color gamut using color-matching functions as defined by the International Commission on Illumination (CIE). The outcomes are presented in the column situated between the simulated (**Figure 2e**) and measured (**Figure 2f**) spectra. Based on the comprehensive spectral data, precise CIE coordinates for color pixels can be calculated using information from Supporting Note 2. From the simulated results, the color gamut could achieve about 161% standard red, green, and blue (sRGB) space and 117% Adobe RGB space. The chromatic coordinates corresponding to the experimentally measured spectra show a reasonable agreement with their simulated counterparts. In the experiment, the color gamut achieved approximately 150% sRGB space and 110% Adobe RGB space. The slightly lower experimental gamut compared to simulations is a consequence of an expanded spectral linewidth, prominently influenced by the numerical aperture (NA) of the objective. The slight deviation arises from the sensitivity of the resonance's spectral position to the angle of incident light, a characteristic inherent in BICs.

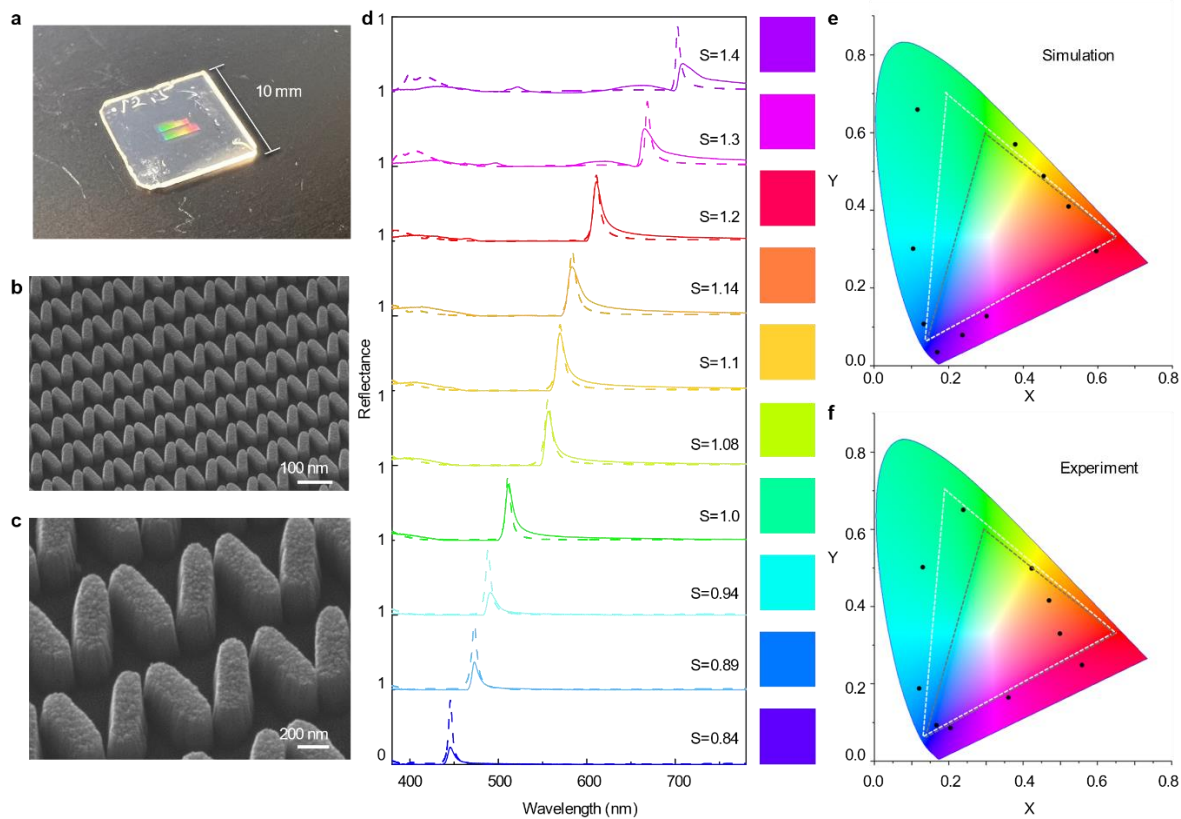

**Figure2. Experimental color hues of the metasurfaces.** (a) Photograph of a manufactured large-scale all-dielectric metasurface, employed in subsequent reflectance measurements. (b,c) SEM micrographs of the metasurface confirm the homogeneity of the nanofabrication. (d) The experimental and simulated spectra depict several rainbow colors observed in distinct samples. The chromaticity coordinates on the CIE 1931 diagram were derived from both simulated spectra (e) and measured spectra (f). The areas enclosed by white and grey dotted lines represent the Adobe RGB and sRGB color spaces, respectively. It is worth noting that both the measured and simulated color spaces exceed the Adobe RGB space.

**Photorealistic color gradients**

Color intensity (brightness) along with chromaticity are vital components of the structural coloration response. The manipulation of reflectance spectrum intensity can be achieved by controlling the q-BIC excitation. As shown in **Figure 3a**, SEM images experimentally verify the accurate reproduction of metasurfaces with asymmetry parameters of θ = 20°, 15°, and 10°. As the asymmetry parameter increases, **Figure 3b** shows a concomitant rise in the reflectance intensity, thereby facilitating the customization of color intensity through precise geometrical asymmetry control. Meanwhile, the reflectance spectra exhibit a noticeable blue shift as observed for increasing geometrical asymmetry, aligning with the numerical predictions. We further optimize the scaling factor of the BIC metasurface at different tilting angles, which remained identical but with a decreasing magnitude. Therefore, we could realize a continuous modification of the color brightness without changes in hue and saturation. Figure S10 shows the measured spectra of the three primary RGB colors, along with continuous intensity before and after scaling factor adjustment.

**Figure 3c** displays photographs of 162 samples featuring distinct scaling factors and tilting angles. It can be observed that the color hue is gradually varied from blue to red with increases of *S* from 0.84 to 1.2. In addition, a smooth color brightness transition can be realized when the tilting angle of the nanoantenna changes from 0 to 20° in steps of 2.5°. In addition, the optical images display a consistent intensity change, while the hue and saturation remain mostly constant. All the results were obtained by photographing the metasurface with an optical microscope (Olympus EP50), and further details can be found in the Methods section.

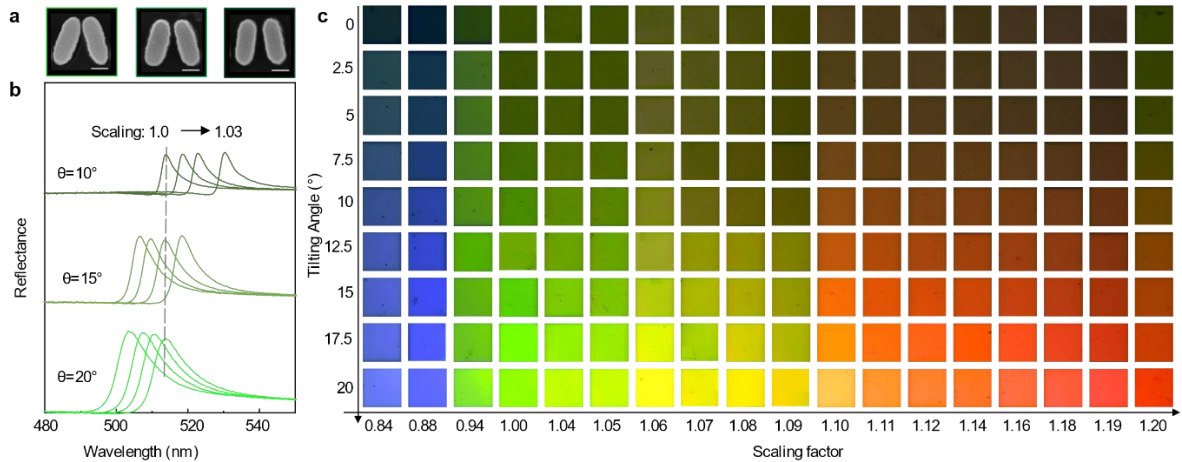

**Figure 3. Continuous color brightness tuning of q-BIC color pixels.** (a) Top-view HR-SEM images show unit cells with varying θ from 10° to 20°. Scale bar: 100 μm. (b) Measured reflectance spectra of the green pixels with the tilting angle varying from 10 to 20°. (c) The color palette was photographed with varying scaling factors and tilting angles of the ellipses. Each sample is 40 μm×40 μm.

**Realistic color image printing with TiO$_2$ metasurfaces**

The manipulation of color brightness while maintaining a constant hue results in a shadow-rendering effect that improves stereo perception. As a proof-of-concept demonstration, a straightforward stereoscopic pattern was chosen as the target image (**Figure 4a**). The experimental photograph of the result is depicted in **Figure 4b**, which highlights the demarcation between the nanostructures tilted at different angles.

By encoding color hue and brightness information into geometrically diverse nanostructures with varying tilting angles, the printed image achieves highly saturated colors and a stereoscopic impression when illuminated with white light. This characteristic enabled the design of a parameter-matching approach to reproduce colored paintings. As illustrated in Figure S11, the procedure involves pixelating the target image, segmenting it into RGB channels, and subsequently associating the color values of each pixel with the closest additive colors present within the metasurface's color palette. The resulting insights are utilized to establish a one-to-one relationship between spectral and spatial information. In accordance with this methodology, **Figure 4c** showcases the simulated image. In **Figure 4d**, the colored metasurface generates a high-resolution image when illuminated with white light under an x-polarizer. The image shows seamless blending of the darker peripheral sides with the black background, accompanied by smooth brightness transitions. Moreover, we explore the influence of various area sizes on the q-BIC resonance, which exhibits an array effect, Figure S12 illustrates that color pixel sizes ranging from 60 μm to 5 μm maintain bright color. Additionally, the study reveals that even with reduced numbers of unit cells (14 × 14, 16 × 16, and 18 × 18) for red, green, and blue colors, the pixels retain their ability to deliver highly saturated colors, which opens up the possibility of full-color stereoscopic printing in the future.

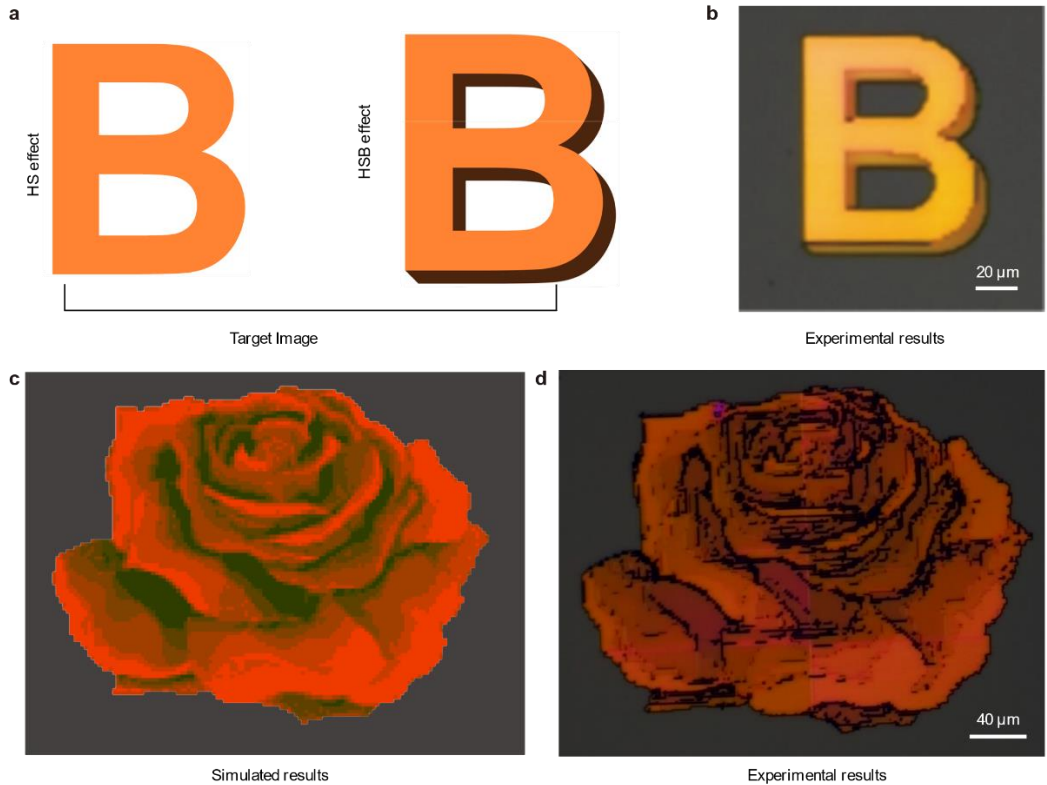

**Figure 4. Photorealistic all dielectric nanopainting.** (a) In comparison to Hue-Saturation (HS) images, HSB (Hue-Saturation-Brightness) images provide a more comprehensive representation of chiaroscuro (i.e., light-dark) information and (b) optical image based on the proposed metasurface. (c) Simulated and (d) experimental results of the encoded color image.

CONCLUSION

We have demonstrated a TiO$_2$ antenna design based on symmetry-protected BICs to realize structural color painting techniques with exceptional color rendition. The q-BIC modes generate ultrasharp reflectance spectra that produce highly saturated color pixels covering the full visible light range. Notably, by varying the geometrical asymmetry of the individual metasurface unit cells, we also achieve full control over the color brightness. The proposed metasurfaces offer an alternative way to obtain individual control over the chromaticity and luminance of the generated colors.

Furthermore, the implementation of numerical optimization strategies effectively facilitated the transformation of square-shaped unit cells into rectangular configurations, providing possibilities for scaled-up fabrication of the proposed metasurfaces, e.g., via deep ultraviolet lithography and nanoimprint lithography[59], to make it possible to accommodate a wide range of photonic and optoelectronic applications[60,61]. Besides, the angle- and polarization-dependent characteristics of the BICs in the presented design offer a promising avenue for the generation of distinct visual effects, particularly polarization-encrypted anti-counterfeiting applications[62,63].

**Methods**

**Numerical simulations**

We obtained the numerical simulation results using the commercial CST Microwave Studio software designed for the THz metasurface and utilizing the time-domain solver that solved Maxwell's equations by finite integral time domain. The numerical simulations of the q-BIC metasurfaces have been performed using the finite element solver contained in CST Microwave Studio, where periodic boundary conditions are applied and the incident light (k) is perpendicular to the metasurfaces plane with x-polarization (TM). Experimentally measured optical constants obtained through ellipsometry were implemented for $TiO_2$ to improve the agreement between simulations and experiments. The refractive indices of $SiO_2$ were assumed to be lossless with n=1.5.

**Nanofabrication**

TiO$_2$ was RF sputtered on SiO$_2$ coverslips at 5 × 10$^{-7}$ Torr at a rate of 0.2 Å/s. Afterward, the positive electron beam resists poly(methyl methacrylate) (950 K, A4, Microresist) was spun onto the sample with soft-baking steps of 3 mins at 180 °C. An electrically conductive polymer (Espacer 300Z) was coated on top of the resist to avoid electron charge accumulations and thus pattern distortions. The lithography pattern was defined using electron beam lithography (Raith eLine plus) with an acceleration voltage of 30 kV, aperture size of 15 μm, a working distance of 10.6 mm, and an area dose of 150 μC/cm$^2$. After the conductive polymer was washed off in a water bath for 10 s, the PMMA layer was developed in a 3:1 isopropanol (IPA): methyl isobutyl ketone solution for 135 s with a subsequent 30 s bath in pure IPA. The metal hard mask consisted of a 50 nm chromium (Cr) layer which was deposited via electron beam evaporation. The lift-off process was conducted in Microposit Remover 1165 overnight at 80 °C, followed by reactive ion dry etching in an RCP-RIE system using an SF$_6$/Ar plasma. Finally, the chrome layer was removed by wet etching with chromium etchant (Sigma-Aldrich)

**Optical Measurement**

The refractive indices and extinction coefficients of TiO$_2$ films were extracted from optical modeling of variable-angle spectroscopic ellipsometry data (J.A. Woollam, M2000XI-210) measured. Ellipsometry spectra were acquired over a range of 210-1690 nm and at four different angles between 65 and 80°.

Reflectance measurements of the fabricated metasurface samples were carried out with a WiTec optical microscope comprising 10× objective (NA = 0.25, Zeiss, Germany) under the illumination of the polarized broadband light source (Olympus TH4-200). The reflectance spectra are normalized to a silver mirror response.

Optical bright-field images are acquired on an Olympus EP50 microscope using a 10× objective (NA = 0.25) and white light from an LED with an x- polarization. For the bright-field images, the white balance is calibrated on a Spectralon Diffuse Reflectance Standard (Labsphere).

ASSOCIATED CONTENT

The **Supporting Information** is available free of charge on the website.

Refractive index of $TiO_2$, optimizing simulated reflectance spectra for maximum color gamut, simulated BIC reflectance spectra with *S* and θ variations, fabrication process, experiment color and its corresponding hue value, unprocessed simulated and measured reflectance spectra, green pixel with different polarization conditions, measured RGB color spectra for tunable color brightness while maintaining hue stability, approach to implementing color mapping for a target image, optical images of various-area $TiO_2$ metasurfaces, color chromaticity calculation and.

AUTHOR INFORMATION

**Corresponding Author**

* E-mail: Andreas.Tittl@physik.uni-muenchen.de

**Notes**

The authors declare no competing financial interest

ACKNOWLEDGMENT

This project was funded by the Deutsche Forschungsgemeinschaft (DFG, German Research Foundation) under grant numbers EXC 2089/1–390776260 (Germany's Excellence Strategy) and TI 1063/1 (Emmy Noether Program), the Bavarian program Solar Energies Go Hybrid (SolTech) and the Center for NanoScience (CeNS). Funded by the European Union (ERC, METANEXT, 101078018). Views and opinions expressed are however those of the author(s) only and do not necessarily reflect those of the European Union or the European Research Council Executive Agency. Neither the European Union nor the granting authority can be held responsible for them. S.A.M. additionally acknowledges the Lee-Lucas Chair in Physics and the EPSRC (EP/W017075/1).

BRIEFS

Table of Contents.

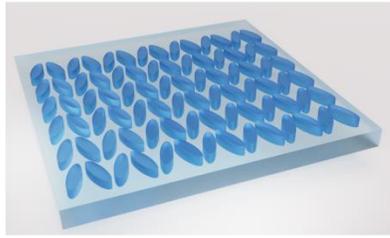
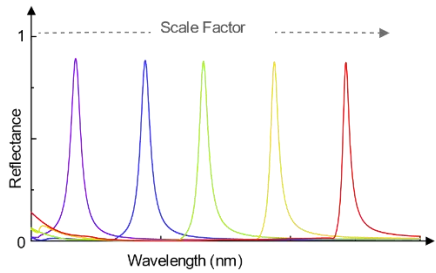
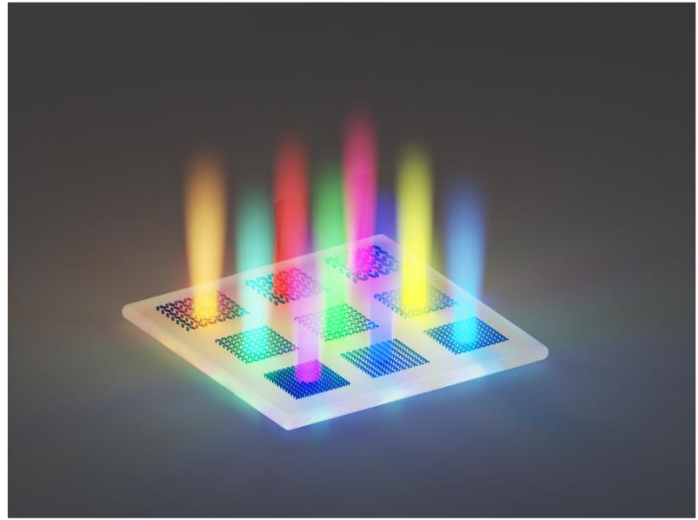

We presented a zig-zag $TiO_2$ metasurface with symmetry-protected BIC, showcasing impressive reflectance, pure color rendition, an extensive gamut palette, and stereoscopic representation.

*Supplementary Material*

**All-Dielectric Structural Coloration Empowered by Bound States in the Continuum**

*Hong Zheng[1,2], Haiyang Hu[2], Thomas Weber[2], Juan Wang[2], Lin Nan[2], Bingsuo Zou[1,3], Stefan A. Maier[4,5] and Andreas Tittl[2]\**

1. Beijing Key Laboratory of Nanophotonics and Ultrafine Optoelectronic Systems, School of Physics, Beijing Institute of Technology, Beijing 100081, China

2. Chair in Hybrid Nanosystems, Nanoinstitute Munich, Faculty of Physics, Ludwig-Maximilians-Universität München, 80539 München, Germany

3. Guangxi Key Lab of Processing for Nonferrous Metals and Featured Materials, School of Resources, Environments and Materials, Guangxi University, Nanning 530004, China

4. School of Physics and Astronomy, Monash University, Clayton, Victoria 3800, Australia.

5. Department of Physics, Imperial College London, London SW72AZ, UK.

\* e-mail: Andreas.Tittl@physik.uni-muenchen.de

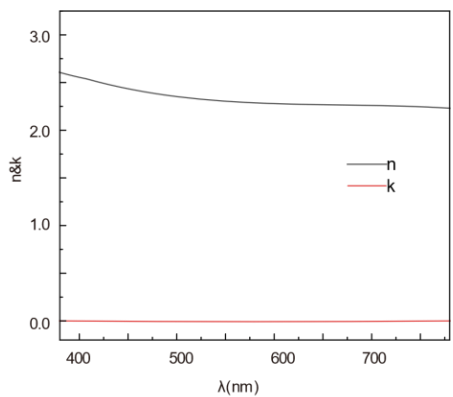

Figure S1. The refractive index (n) and extinction coefficient (k) for the amorphous $TiO_2$ films.

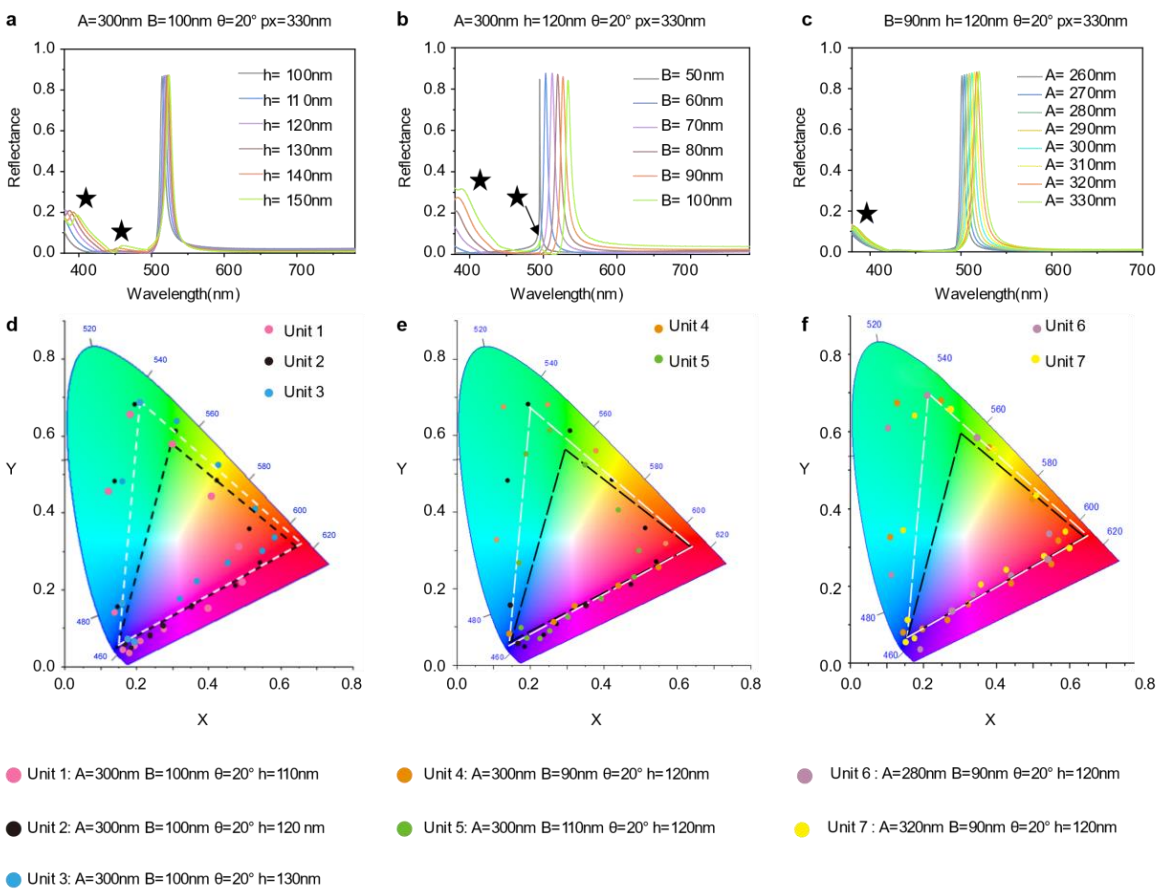

Figure S2. (a-c) The simulated reflectance spectra versus the dimensions of the nanoresonators with periodic $P_x$ fixed to 330 nm. (d-f) Chromatic coordinates in the CIE 1931 diagram obtained from the simulated spectra.

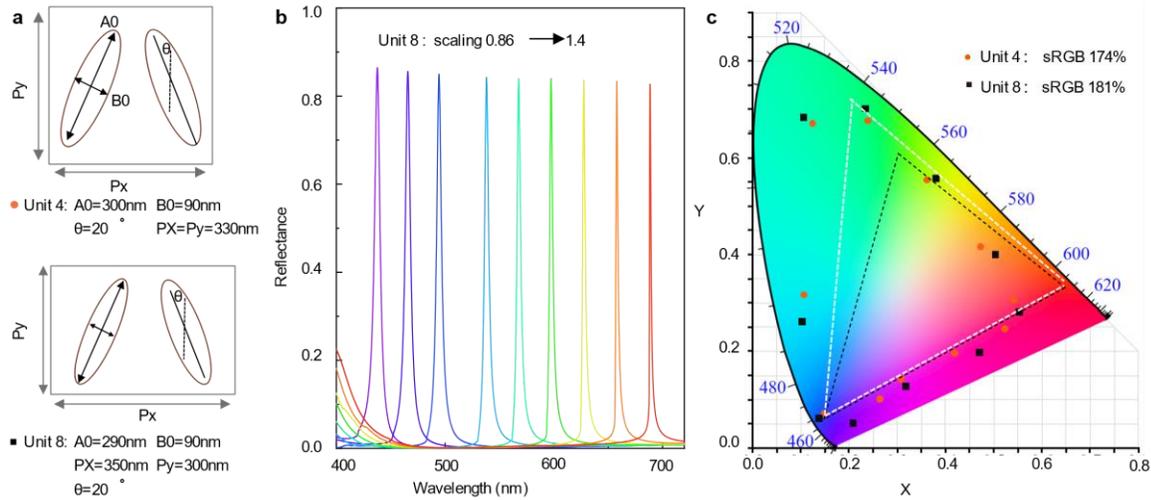

Figure S3. Further optimization of color gamut by introducing nanoresonators with various Px & Py.

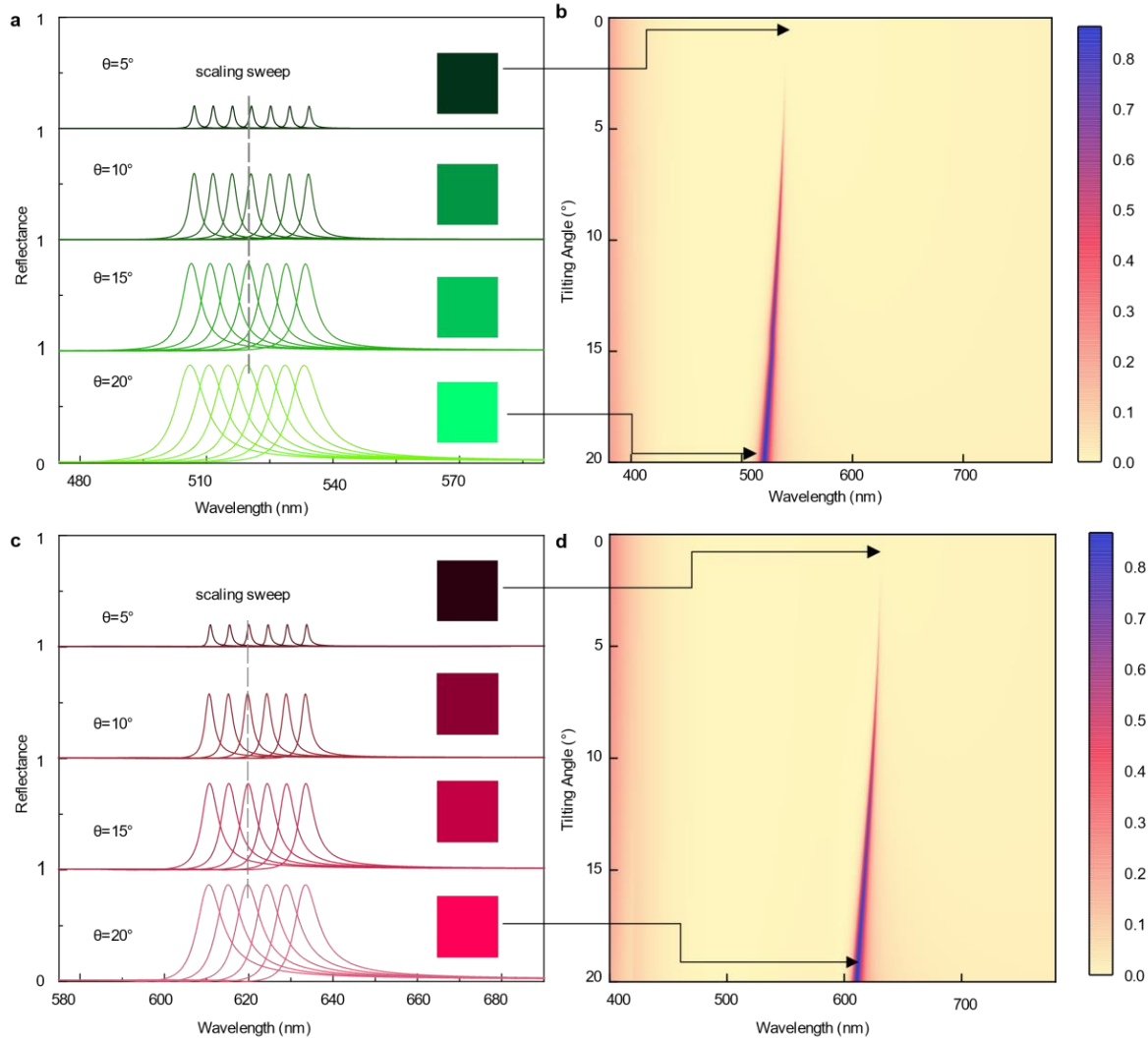

Figure S4. The simulated reflectance spectra of BICs show variations in scaling factors (S) and asymmetries (θ). The scaling factor S is kept fixed at 1.0 (a) and 1.2 (c). The insets display the corresponding calculated colors. The reflectance spectra are analyzed for different tilting angles of the unit cell, while maintaining the scaling factor S fixed at 1 (b) and 1.2 (d).

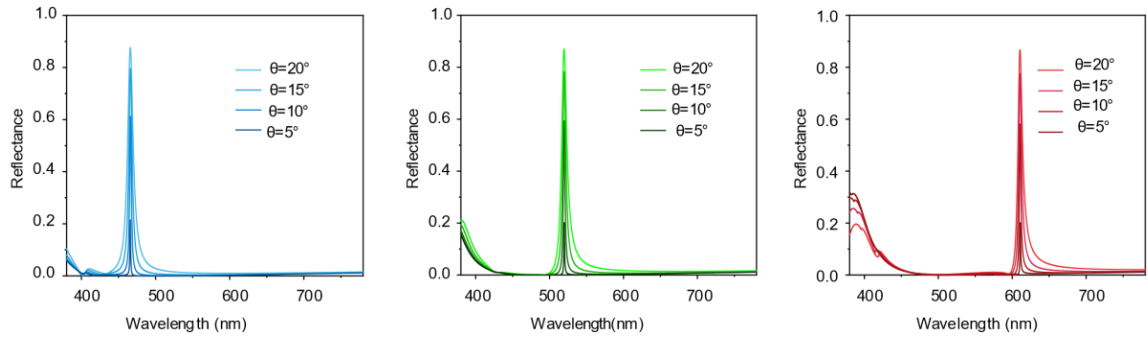

Figure S5. Simulated reflectance spectra of the blue, green and red pixels with the tilting angles of the nanoapertures varying from 0 to 20° after scaling sweep from Figure 1d, Figure S4a and S4c.

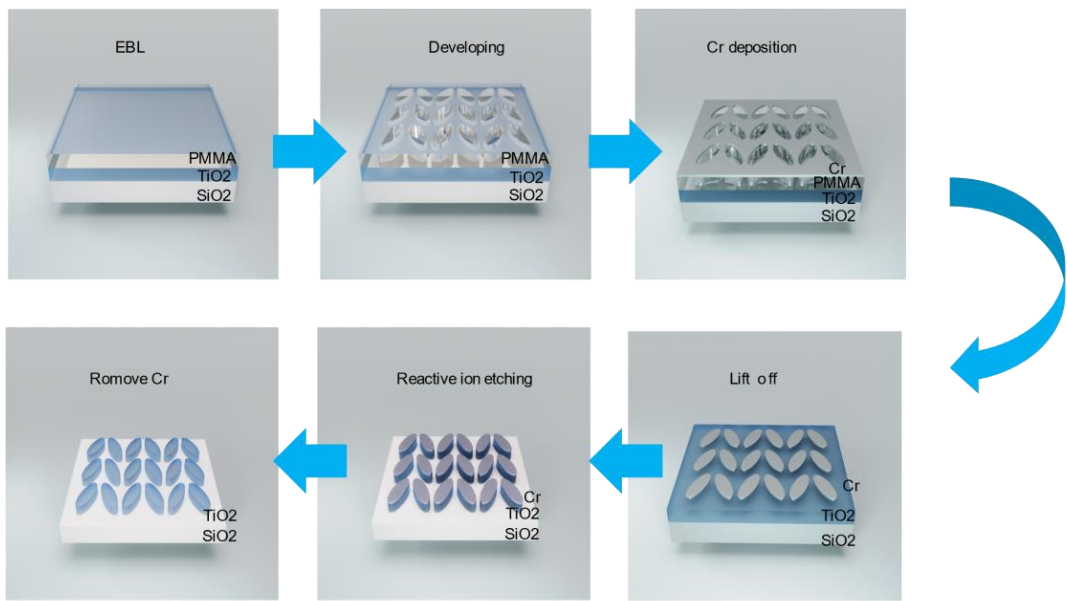

Figure S6. The schematic of the fabrication process for $TiO_2$ metasurafce.

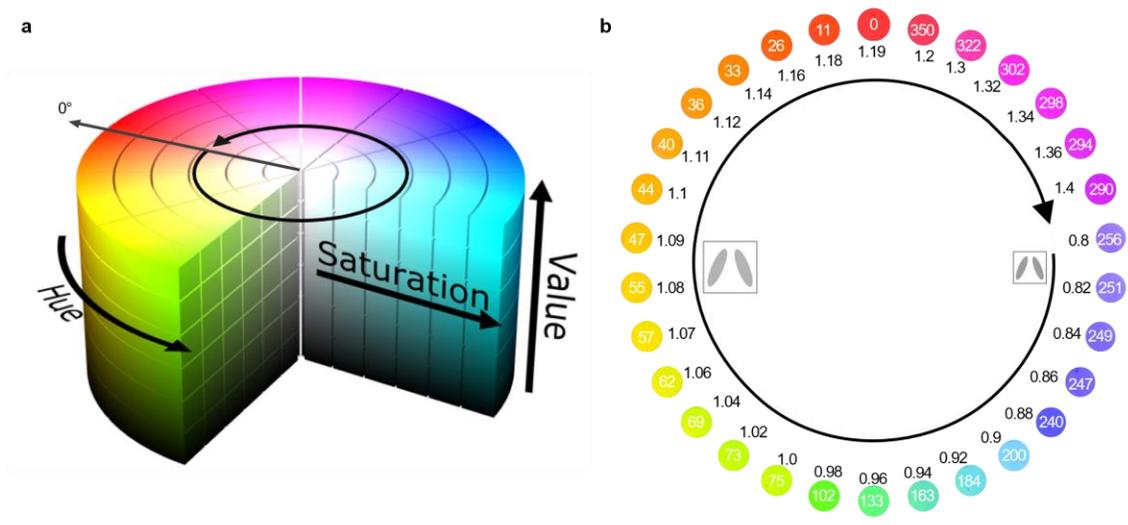

Figure S7. Experimental colors and their corresponding hue values in the HSV color space.

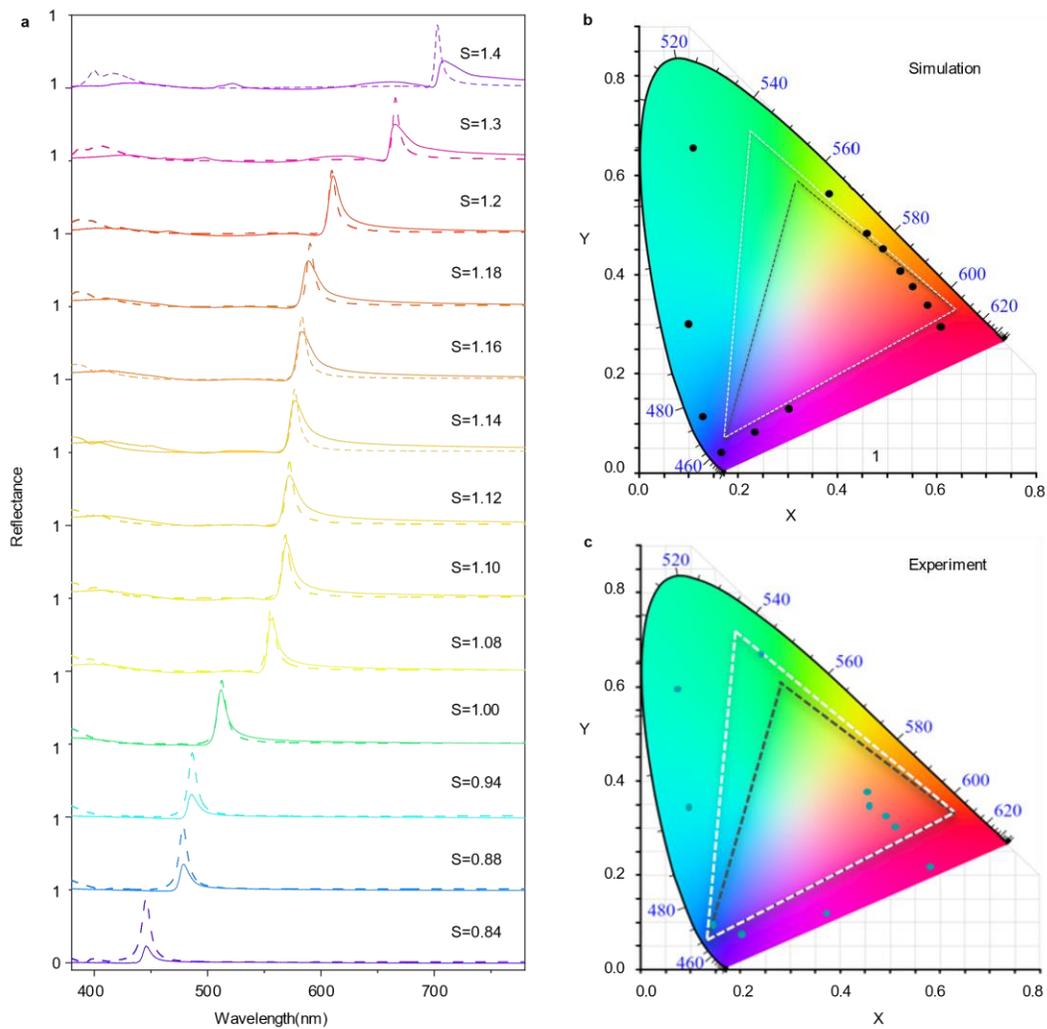

Figure S8 (a) Unprocessed simulated and measured reflectance spectra. (b) CIE 1931 chromaticity coordinates with different colors obtained in the simulations and experiments.

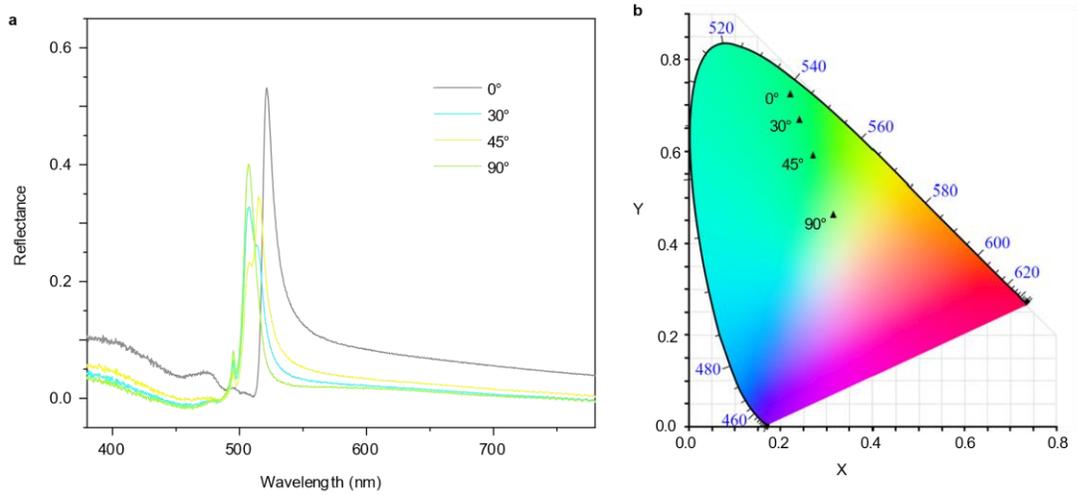

Figure S9. Characterization of the fabricated green pixel with different polarization conditions.

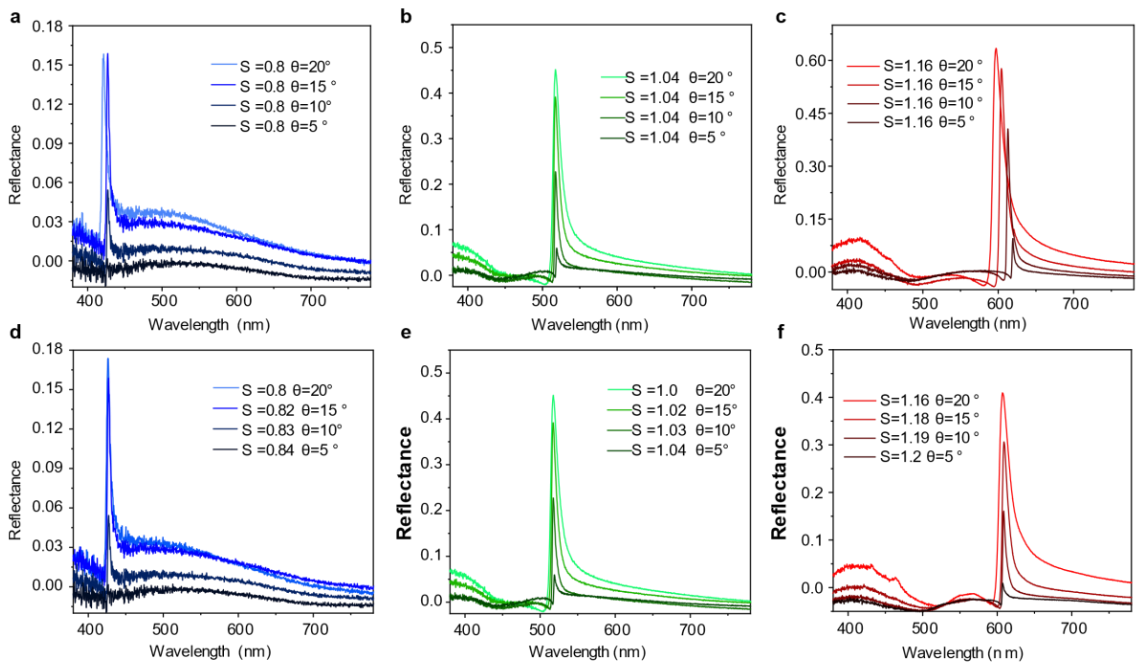

Figure S10. Measured spectra of the three primary RGB colors, along with continuous intensity before (a-b) and after(c-f) scaling factor adjustment.

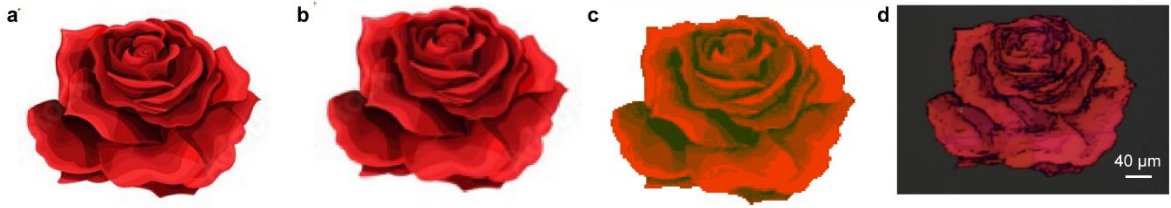

Figure S11. A schematic diagram illustrating the design of a patterned metasurface to achieve diverse color hues and brightness simultaneously. (a) A target image. (b) Utilizing the pixelation process to resize the target image to a desired dimension. (c) The pixilated image in settings from the database. (d) Optical experimental micrograph.

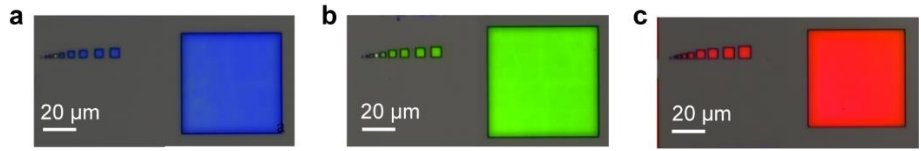

Figure S12. Optical images of (a) red, (b)green, and (c)blue TiO$_2$ metasurfaces of different areas, the lateral size changes from 60 to 5 μm.

Supporting Note 1: Determination of reflectance equations

The far-field reflectance can then be calculated according to temporal coupled mode theory (TCMT) approach[1,2], which are associated with the radiative $\gamma_{rad}$ and intrinsic $\gamma_{int}$ losses, respectively. The reflectance R can be expressed as:

$$R = \left| \frac{\gamma_{rad}}{\gamma_{int} + \gamma_{rad} + i(\omega - \omega_0)} \right|^2$$

And the relationship between $\gamma_{rad}$ and asymmetry factor $\alpha$ follows the characteristic quadratic relationship:

$$\gamma_{rad} \sim \alpha^2 = \sin^2 \theta$$

Supporting Note 2: Color chromaticity calculation

The experimental spectra obtained from color pixels can be rsformed into their respective CIE coordinates by employing the following integrals:

$$X = \int_{380}^{780} I(\lambda)\, S(\lambda)\, \bar{x}(\lambda)\, d\lambda$$

$$Y = \int_{380}^{780} I(\lambda)\, S(\lambda)\, \bar{y}(\lambda)\, d\lambda$$

$$Z = \int_{380}^{780} I(\lambda)\, S(\lambda)\, \bar{z}(\lambda)\, d\lambda$$

where $I(\lambda)$ is the spectral power distribution of the illuminant, $S(\lambda)$ is the measured spectrum and $\bar{x}(\lambda)\ \bar{y}(\lambda)\ \bar{z}(\lambda)$ are the red, green and blue color matching functions. The integrals are computed across the entire visible wavelength spectrum, ranging from 380 nm to 780 nm.

Subsequently, the X and Y values were normalized to derive the corresponding CIE (x, y) coordinates:

$$x = \frac{X}{X + Y + Z}$$

$$y = \frac{Y}{X + Y + X}$$

Finally, the data can be visualized on the CIE 1931 color diagram, where the luminance of a color is directly associated with its Y coordinate.